# Effect of isotopic mass on the photoluminescence spectra of β zinc sulfide

F. J. Manjón[1*], M. Mollar[1], B. Marí[1], N. Garro[2], A. Cantarero[2], R. Lauck[3], and M. Cardona[3]

[1] Dpto. de Física Aplicada, Universitat Politècnica de València, C/. Vera s/n, 46012 València (Spain)

[2] Institut de Ciència dels Materials, Universitat de València, P.O. Box 22085, E-46071 València (Spain)

[3] Max-Planck-Institut für Festkörperforschung, Heisenbergstr. 1, D-70569 Stuttgart (Germany)

**Abstract:** Zinc sulfide is a wide bandgap semiconductor which crystallizes in either the wurtzite modification (α-ZnS), the zincblende modification (β–ZnS) or as one of several similar tetrahedrally coordinated polytypes. In this work, we report a photoluminescence study of different samples of isotopically pure β-ZnS crystals, and crystals with the natural isotopic abundances, at 15 and 77 K. The derivatives of the free and bound exciton energies on isotopic mass have been obtained. They allow us to estimate the contribution of the zinc and sulfur vibrations to the bandgap renormalization energy by electron-phonon interaction. A two-oscillator model based on the zinc and sulfur renormalization energies has been used to account for the temperature dependence of the bandgap energy in ZnS. The results are compared with those found for other tetrahedrally coordinated semiconductors.

---

* Corresponding author. E-mail address: fjmanjon@fis.upv.es
  Phone: +34 96 652 84 42, Fax:+34 96 652 84 09

# I. INTRODUCTION

Zinc sulfide (ZnS) is a widely used material for electroluminescent devices and is also suited for field emission displays [1]. It is also a promising wide-bandgap material for the development of LEDs and lasers operating in the UV range since it exhibits a fundamental direct bandgap energy of 3.84 eV (at low temperature, 3.72 eV at room temperature) [2].

This semiconductor is found in nature both in the hexagonal wurtzite ($\alpha$-ZnS) and the cubic zincblende ($\beta$-ZnS) modifications. The $\beta$ modification is also called "sphalerite"; together with zincblende this name is employed to designate its generic cubic crystal structure. ZnS, both synthetic and natural, also crystallizes in as many as ten different polytypes [3] transforming from one into the other at stacking faults [4] Unintentionally-doped ZnS is n-type and usually exhibits a large free electron concentration due to the presence of unbalanced donor- and acceptor-like defects, often impurities, which are responsible for the different colors, from pale yellow to dark green, shown by bulk ZnS crystals. This problem is common to many strongly polar semiconductors such as ZnO and GaN.

The natural isotopic abundance of sulfur is 95% $^{32}$S, 0.5% $^{33}$S, 4.2% $^{34}$S; and 0.02% $^{36}$S. Within the accuracy of the experiments reported here it will be assumed, for simplicity, that $^{nat}$S is pure $^{32}$S. The natural abundance of Zn is 48.6% $^{64}$Zn, 27.5% $^{66}$Zn, 4.1% $^{67}$Zn, 18.8% $^{68}$Zn, and 0.6% $^{70}$Zn. Therefore, in the spirit of the virtual crystal approximation (VCA) implicitly assumed here, the average natural composition of ZnS ($^{nat}$Zn$^{nat}$S) will be taken to be $^{65.4}$Zn$^{32}$S.

It is well known that the semiconductor bandgap energy depends on the average isotopic mass [5,6]. This dependence has been studied theoretically for $\beta$-ZnS and other zincblende-type semiconductors [7]. In previous work, the bandgap shifts caused by the change in the isotopic composition of Zn and O in ZnO and of N in GaN (both wurtzite-type) have been estimated from photoluminescence (PL) and/or cathodoluminescence measurements [8,9]. In the present work, we report a study of the low-temperature PL signal of several unintentionally-doped β-ZnS samples with different isotopic compositions. Our main interest is the investigation of the shift of the PL features with the isotopic masses in order to obtain the isotopic contributions of zinc and sulfur to the electron-phonon renormalization of the bandgap energy and to model the temperature dependence of this energy.

**II. EXPERIMENTAL PROCEDURE**

The zincblende ZnS single crystals used in this study were grown by chemical vapor transport in a silica ampoule where a porous layer of zinc is formed after sublimation of zinc (99.9995%). The β-ZnS synthesis takes place in the ampoule under sulfur vapor using iodine as a transport agent. More details are given elsewhere [10]. Four batches of β-ZnS crystals with pure isotopic composition of Zn ($^{64}$Zn$^{32}$S, $^{68}$Zn$^{32}$S, $^{64}$Zn$^{34}$S, and $^{68}$Zn$^{32}$S$_{0.5}$$^{34}$S$_{0.5}$) and one with the natural isotopic composition of Zn ($^{nat}$Zn$^{32}$S) were investigated. Unpolarized low-temperature PL experiments were performed at 15 and 77 K with the five samples in a backscattering geometry: Because of the high Debye temperature of ZnS (440 K), both 15 K and 77 K correspond to the low-temperature limit. For the PL measurements at 77 K, the samples were placed inside a liquid nitrogen cryostat and were excited with the 275.4 nm (4.5 eV) line of an Ar$^+$-ion laser at power levels of 20 mW (100 W/cm$^2$ excitation density at the sample). The emitted light was analyzed by a Jobin Yvon T64000 triple spectrometer using a



CCD detector. For the PL measurements at 15 K, the samples were placed inside a helium close-cycle cryostat and were excited with the 325 nm (3.81 eV) line of a He-Cd laser at power levels of 30 mW (10 W/cm$^2$ excitation density at the sample). The emitted light was analyzed by a TRIAX 550 spectrometer using either a GaAs PMT or a CCD detector optimized for the UV-VIS range.

**III. RESULTS AND DISCUSSION.**

**Figure 1(a)** shows the near-bandgap PL spectra of a $^{68}$Zn$^{nat}$S sample at 77 K excited above the bandgap. The spectra are dominated in most of the samples studied by two major bands. One band is located at 326.8 nm (3.794 eV) and can be assigned to the emission of the free exciton. The other band is centered at 327.6 nm (3.785 eV) and can be attributed to an exciton bound to a neutral donor [2]. **Figure 1(b)** shows representative PL spectra of the $^{64}$Zn$^{32}$S, $^{68}$Zn$^{32}$S, $^{64}$Zn$^{34}$S and $^{nat}$Zn$^{32}$S samples at 77 K in the excitonic region of **Fig 1(a).** The crystalline quality of the samples investigated in this study was rather inhomogeneous. Consequently, there is a scatter in the energies and linewidths of the free and bound exciton peaks measured in different points of the same sample of typically 0.5meV. Thus, the excitonic energies for each isotopic composition are obtained by averaging over several spectra.

In order to reduce the experimental scatter, we have used in our analysis the results obtained from samples that exhibited the narrowest peaks in the PL spectra. We have found that most of the average values agree with those of the optimal samples; therefore, we have plotted in **Fig. 1(b)** the PL bands of samples whose bound exciton energies are close to the average value for each isotopic composition. A clear blue shift



of the energy of the free and bound exciton PL band is observed in all spectra when increasing either of the isotopic masses.

**Figure 2(a)** shows the dependence of the average peak energy of the free exciton at 77 K versus the total mass of the ZnS molecule. Similarly, **Fig. 2(b)** shows the dependence of the average peak energy of the bound exciton at 77 K versus the total mass of the ZnS molecule. From the observed average energies we have obtained isotopic mass coefficients for the energy shift of the free and bound excitons with increasing isotopic mass ($dE_x/dM_{a,c}$, where the subindices a and c refer to the derivative with respect to the anion or the cation mass, respectively). The different isotopic mass coefficients of the free and bound exciton energy are obtained from the slopes of the lines in **Figs. 2(a) and 2(b)** and given in parenthesis in those figures.

For the free exciton energy shift, we have obtained a zinc isotopic coefficient $dE_x/dM_{Zn} = 0.31 \pm 0.20$ meV/amu when the Zn mass is changed and the sulfur mass remains equal to 32 (notice the large error bars). A similar Zn isotopic mass coefficient for a constant sulfur mass equal to 32, but with smaller error bars, has been obtained from the shifts of the bound exciton energy ($0.40 \pm 0.02$ meV/amu). These isotopic mass coefficients are similar to that found previously for the change of the A free exciton with the Zn mass in ZnO ($0.42 \pm 0.03$ meV/amu) [8]. They are somewhat larger than the corresponding coefficient measured and calculated for the direct gap of germanium (after dividing by two to take into account the number of atoms per unit cell): 0.25 meV/amu [7].

The isotopic mass coefficient of sulfur, obtained from the free exciton energy shifts when the S mass is changed and the zinc mass remains equal to 64, is $dE_x/dM_S = 0.79 \pm 0.05$ meV/amu. This value is considerably smaller than that obtained from the



bound exciton energy shifts (1.22 ± 0.12 meV/amu) under the same conditions. Unfortunately, we could not find exciton-related PL signals for $^{68}Zn^{32}S_{0.5}^{34}S_{0.5}$ samples at 77 K, hence we have no estimates of the S isotopic mass coefficient obtained for zinc masses equal to 68.

In order to obtain other estimates of the S isotopic mass coefficient and to try to understand the large deviation of the sulfur isotopic mass coefficients derived from the two types of excitons, we have analyzed other bands of the spectra appearing at lower energy at 15 K under He-Cd laser excitation (3.81eV). **Figure 3** shows the typical spectrum of some ZnS samples in the 3.40-3.75 eV region at 15 K. In this energy range the spectra of some samples is dominated by a band around 3.690 eV and donor-acceptor pair (DAP) transitions with LO phonon-related replicas whose zero-order peak is around 3.626 eV [11]. Our spectra in this region resemble those found in natural ZnS doped with I and Na [11,12]. Correspondingly, we have interpreted our 3.690 eV band as arising from a free-to-bound acceptor $(e,A^0)$ transition due to its similarity to that found at 3.677 eV in ZnS doped with I and Na measured at 4.2 K [2,11,12]. At present we do not know the nature of the acceptor involved, its binding energy is 13 meV smaller than that reported for a Na acceptor since both the free-to-bound acceptor band and the zero-order DAP peak are shifted by the same amount with respect to previous measurements [11,12]. In previous works, the Na acceptor energy was estimated to be 178 meV; in our samples the same procedure yields an acceptor energy of 165 meV. From the free-to-acceptor and zero-order DAP peak we find a zinc isotopic mass coefficient $dE_g/dM_{Zn}$ = 0.44 ± 0.04 meV/amu. On the other hand, we have found the sulfur isotopic mass coefficient to be $dE_g/dM_S$ = 1.10 ± 0.10 meV/amu using the DAP peak and $dE_g/dM_S$ = 1.70 ± 0.20 meV/amu using the free-to-acceptor $(e,A^0)$ peak.



From the above measurements we conclude that the best value of the zinc isotopic mass coefficient of the bandgap is obtained from the bound exciton energy ($dE_g/dM_{Zn}$ = 0.40 ± 0.02 meV/amu) while the best value of the sulfur isotopic mass coefficient is obtained from the free exciton energy ($dE_g/dM_S$ = 0.79 ± 0.05 meV/amu). This zinc isotopic mass coefficient agrees with the value found previously for ZnO [8]. The value of the sulfur isotopic mass coefficient also agrees with the sulfur isotopic coefficient of 0.71 meV/amu recently obtained for wurtzite-type CdS [13] and also with 0.72 and 0.95 meV/amu found previously for CdS [14]. Furthermore, our value is consistent with the somewhat larger value found for the P isotopic mass coefficient of the bandgap in GaP (0.95 meV/amu) [6].

We believe that the large values of the sulfur isotopic mass coefficient obtained from the bound exciton, free-to-acceptor and zero-order DAP peak energies are likely to be related to the presence of several shallow donors in the samples, as it occurs in ZnO, where native defects and H impurities are known to contribute as shallow donors to the n-type conductivity [15]. Another possibility in the case of the bound excitons is that for ZnS the different radiative emission lines due to the shallow donor-to-free hole recombination ($D^0$,h) and the shallow donor bound exciton ($D^0$,X) are not expected to be separately resolved, like in the case of CdTe [16]. It should be recalled that the isotopic coefficients just discussed include not only electron-phonon effects but also the effect of the anharmonic renormalization of the lattice parameter (thermal expansion) [5].

The renormalization of the bandgap energy $\Delta E_g$ due to isotopic substitution is proportional to $M^{-1/2}$ [5]; therefore, the difference between the zero-point bandgap



renormalization energies for two different isotopic compositions *A* and *B* can be described by the linearized relation

$$\Delta E_g(A) - \Delta E_g(B) = \Delta E_g \left( \frac{M(A) - M(B)}{M(A) + M(B)} \right), \qquad (1)$$

where $\Delta E_g$ *(A)* and $\Delta E_g$ *(B)* represent the zero-point gap renormalization for the two compositions *A* and *B* with masses *M(A)* and *M(B),* respectively. Equation (1) allows us to estimate the contribution of each of the two constituents of a binary compound to the renormalization $\Delta E_g$ from the corresponding measured isotope shift $\Delta E_g$ *(A)* - $\Delta E_g$ *(B)*. Because of the small exciton binding energy, the band edge shifts can be assumed to equal the corresponding shifts of the free excitons. The same applies to bound excitons provided the defects the excitons are bound to are the same for sample *A* and for sample *B*. Taking the more accurate Zn isotopic mass coefficient of the donor bound exciton (0.40 ± 0.02 meV/amu) we obtain with Eq.(1) a contribution of the zero-point vibrations of zinc to the gap renormalization of β-ZnS $\Delta E_g$ (Zn)= –53 ± 4 meV and the corresponding contribution of sulfur $\Delta E_g$ (S)= –52 ± 4 meV. Adding these contributions, a total zero point gap renormalization energy $\Delta E_g$ = –105 ± 7 meV is found.

In order to compare these gap renormalizations with those obtained from fits to the temperature dependence of the bandgap energy, we have plotted in **Fig. 4** the corresponding dependence measured for the free exciton in natural ZnS thin films up to 550 K [17], and fitted it to a two-harmonic-oscillator model [5,8,9,18]. In this model, the temperature dependence of the bandgap effected by the electron-phonon interaction (plus the thermal expansion contribution) is described by the expression:



$$E(T) = E_0 - 2M_{Zn}\frac{dE}{dM_{Zn}}\left[\frac{2}{\exp(\theta_1/T)-1}+1\right] - 2M_S\frac{dE}{dM_S}\left[\frac{2}{\exp(\theta_2/T)-1}+1\right] \quad (2)$$

where $E_0$ is the bare bandgap energy, $dE_g/dM_{Zn}$ and $dE_g/dM_S$ are the zinc and oxygen isotopic coefficients of the bandgap, and $\theta_1$ and $\theta_2$ are the average temperatures of the zinc and sulfur vibrations which, because of the difference between the masses of these two elements, will be assumed to correspond to acoustic and optical phonons, respectively [10].

Because of the large error bars which affect the measured isotopic coefficients (mass derivatives in **Eq.(2)**) a fit to the experimental points of **Fig. 4** setting the isotope coefficients equal to those determined above and varying only $\theta_1$ and $\theta_2$ does not lead to good results. Instead we have tried to vary, beside $\theta_1$ and $\theta_2$, also the isotope mass coefficients but four parameters are too many to fit the limited experimental information contained in **Fig. 4**. We have, therefore, fixed $E_g(0)$ = 3.807 eV and $\theta_1$ = 215 K (148 cm$^{-1}$) (this value of $\theta_1$ corresponds to the middle of the acoustic phonon bands [19]), and obtained the excellent fit shown by the solid curve in **Fig. 4** with the parameters: $dE_g/dM_{Zn}$ = 0.31 meV/amu, $dE_g/dM_S$ = 0.59 meV/amu, and $\theta_2$ = 455 K (313 cm$^{-1}$). This latter value corresponds to an average optic phonon that lies in the middle of the TO phonon band [19].

The value of the zinc isotopic mass coefficients found from this fit agrees with that measured for the free exciton whereas that for sulfur is somewhat smaller than 0.79 meV/amu extracted from **Fig. 2a**, a fact which may have a number of causes. It may be due to the relatively small temperature range of the measurements in **Fig. 4** (T<550K) and also to the well-known failure of the two oscillator model which does not reflect the $T^4$ dependence of the gap shift obtained at low temperature [20]. Three oscillator fits correct in part this deficiency [21] but the number of free parameters needed is again too large to yield meaningful fit parameters. The discrepancy may also be related, at least in part, to the large error bars affecting the determination of the isotopic mass coefficient. The coefficient obtained from the measured shift in the free exciton when varying $M_{Zn}$, $dE_g/dM_{Zn} = (0.31 \pm 0.2)$, agrees with the one obtained from the fit in **Fig. 4** (also 0.31 meV/amu) agrees within the experimental error. *A priori* the free exciton coefficient should be more reliable than that obtained from bound excitons ($dE_g/dM_{Zn} = (0.31 \pm 0.2$ meV/amu) even though the error bars of the latter are smaller. We should also mention that the values of the isotopic mass coefficients calculated semiempirically in [7] are $dE_g/dM_{Zn} = 0.44$ meV/amu and $dE_g/dM_S = 1.28$ meV/amu. These values represent only the electron-phonon interaction effect and must be increased by the thermal expansion contribution which is not known. Taking for this contribution the values calculated for ZnSe [7] the enhancement would be 0.12 meV/amu for zinc and 0.04 meV/amu for sulfur.

The dashed line in **Fig. 4** represents the asymptotic behavior found from the fit with **Eq.(2)** at high temperatures, extrapolated to $T=0$. The intercept with the vertical axis yields an estimate for the bare (unrenormalized) gap and the difference between this intercept and $E_g(0) = 3.807$eV, of about -80 meV. This value represents the total zero point renormalization which from the measured isotopic mass coefficients was estimated to be $-105 \pm 7$ meV. The larger value found from the latter estimate reflects the already mentioned fact that the mass coefficients measured directly are somewhat larger than those obtained from the fit. A value of -80 meV has also been found by Pässler [18] by means of a two-oscillator fit to the data of **Fig. 4**. Differences in the gap renormalizations obtained by the two procedures discussed above have also been found in a recent paper on GaN [9] in spite of the fact that the temperature range of the gap measurements extended to 1100 K. In any case, a temperature range wider than that of **Fig. 4** would be desirable in order to obtain more accurate fit parameters and gap renormalizations.

Before closing, a few comments about the meaning of the fitted values of $\theta_1$ and $\theta_2$ are in order. As already mentioned, the value $\theta_1=148$ cm$^{-1}$, kept fixed during the fit, falls in the middle of the acoustic bands (TA and LA) which extend from 0 to 200 cm$^{-1}$ [19]. The frequency $\theta_2 = 350$ cm$^{-1}$ lies in the middle of the TO band which has twice as much weight as the LO band centered at 320 cm$^{-1}$ [19].

## IV. CONCLUSIONS.

We have reported a detailed study of the low-temperature PL of several unintentionally-doped β-ZnS samples with different isotopic compositions. As a result, we have obtained the zinc and sulfur isotopic mass coefficients of the bandgap of β-ZnS from the isotopic shift of the free and bound exciton energies. The zero-temperature total bandgap renormalization energy in β-ZnS has been estimated from the isotopic mass coefficients of the bandgap. This value has been found to be somewhat larger than that obtained from fits of the temperature dependence of the free exciton energy, measured up to 550 K, with a two-oscillator fit using three free parameters and also larger than that obtained from the linear extrapolation of the high-temperature data for $E_g$ *(T)*. We believe that bandgap measurements above 550 K are needed in order to perform more reliable fits and extrapolations. The results obtained are similar to those found for related tetrahedral semiconductors such as germanium and GaP. The mass coefficient of the anion is found to be larger in ZnO and GaN, a fact which is common to atoms of the same row of the periodic table, including diamond [5].

**Acknowledgments:** The authors thank Reinhard Kremer for a critical reading of the manuscript. This work was supported by a Spanish Government MCYT grant MAT2002-04539-C02-02 and Generalitat Valenciana OCYT grant GV01-211 and also by the Fonds der Chemischen Industrie.

# Figure captions

**Fig. 1.** (a) Near-bandgap PL spectra of a $^{68}$Zn$^{32}$S bulk sample at 77 K excited with a laser operating at 275.4 nm. The two peaks observed have been assigned to free (FE) and bound (BE) exciton recombination (b) Selected photoluminescence spectra of ZnS samples with isotopic composition at 77 K in the FE and BE range.

**Fig. 2.** (a) Average peak energy of the free exciton at 77 K as a function of the total mass of the ZnS molecule. A straight line connects the $^{64}$Zn$^{32}$S and $^{64}$Zn$^{32}$S points; its slope determines the sulfur isotopic mass coefficient. The other straight line represents the least squares fit to the three points used to estimate the zinc coefficient: note the large error bars . (b) Average peak energy of the bound exciton at 77 K as a function of the total mass of the ZnS molecule. The least squares fit to the data for $^{32}$S give the zinc mass coefficient whereas a line through only two points determines the sulfur mass coefficient. The numbers in parenthesis are the corresponding zinc and sulfur isotopic mass coefficients (in meV/amu).

**Fig. 3.** Representative spectrum of a natural ZnS sample in the 330-370 nm region at 15 K when excited with a laser operating at 325 nm (3.81 eV). A free-to-bound acceptor peak and a zero-phonon DAP with its LO phonon replicas are clearly seen.

**Fig. 4.** Energy of the free exciton in natural β-ZnS thin films for temperatures up to 550 K, as obtained from Ref. [17]. The solid curve corresponds to a fit using Eq. (2) and adjusting three parameters (see text). The dashed straight line represents the asymptotic behavior for T $\rightarrow$ ∞ according to that fit. Its intercept with the vertical axis gives the unrenormalized (bare) value of the gap.

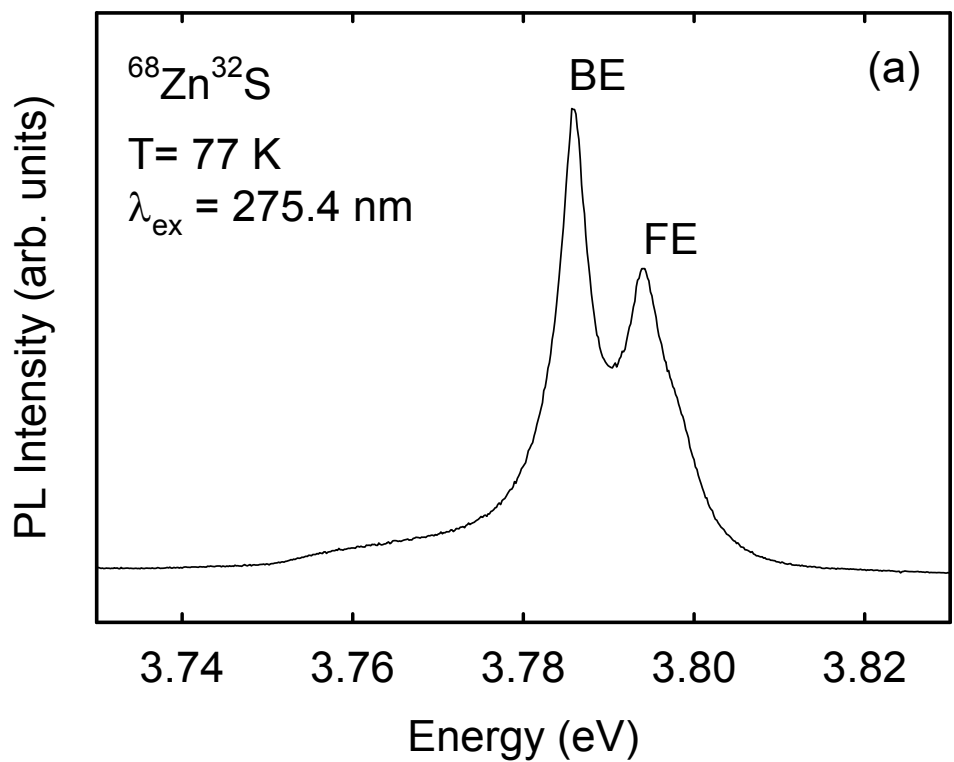

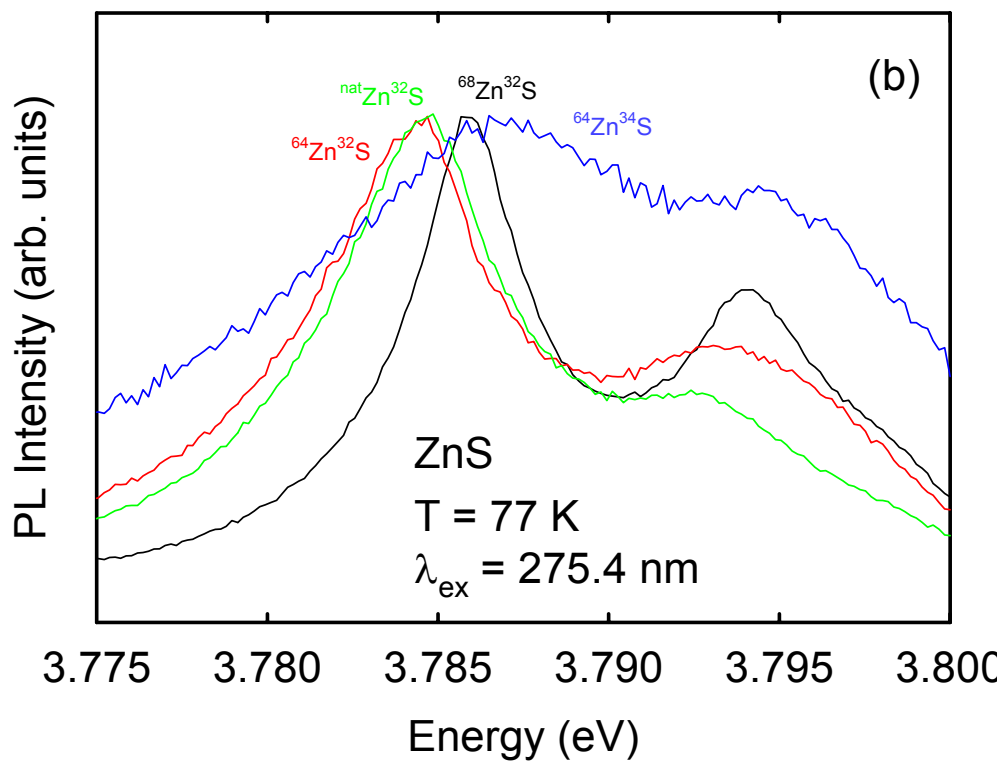

Fig. 1. Manjón

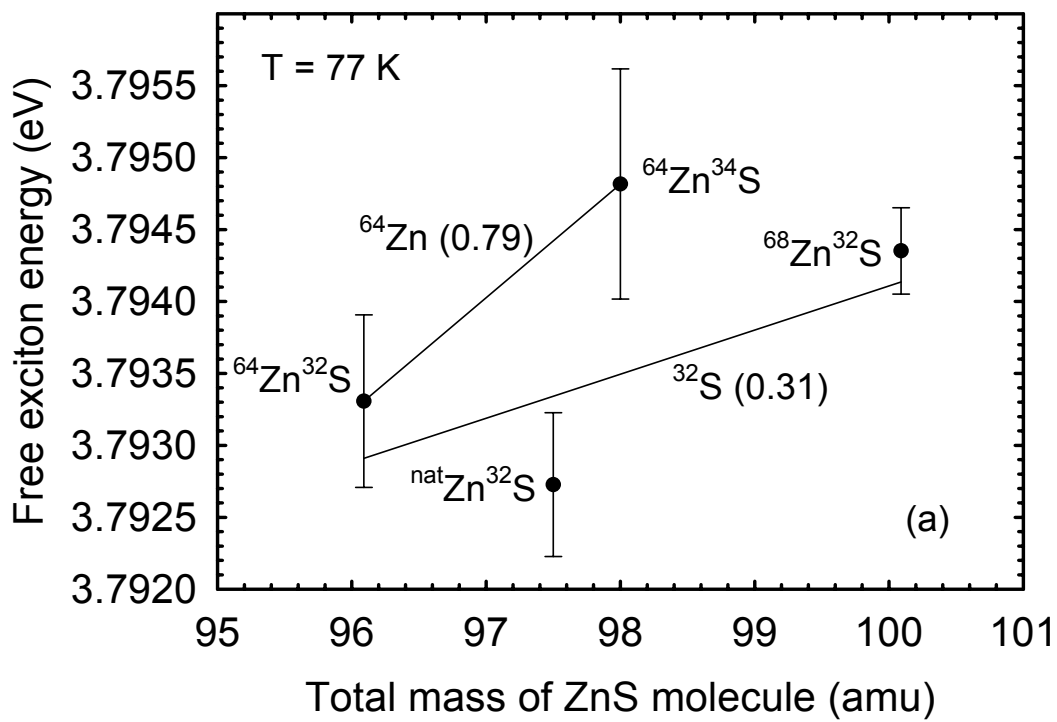

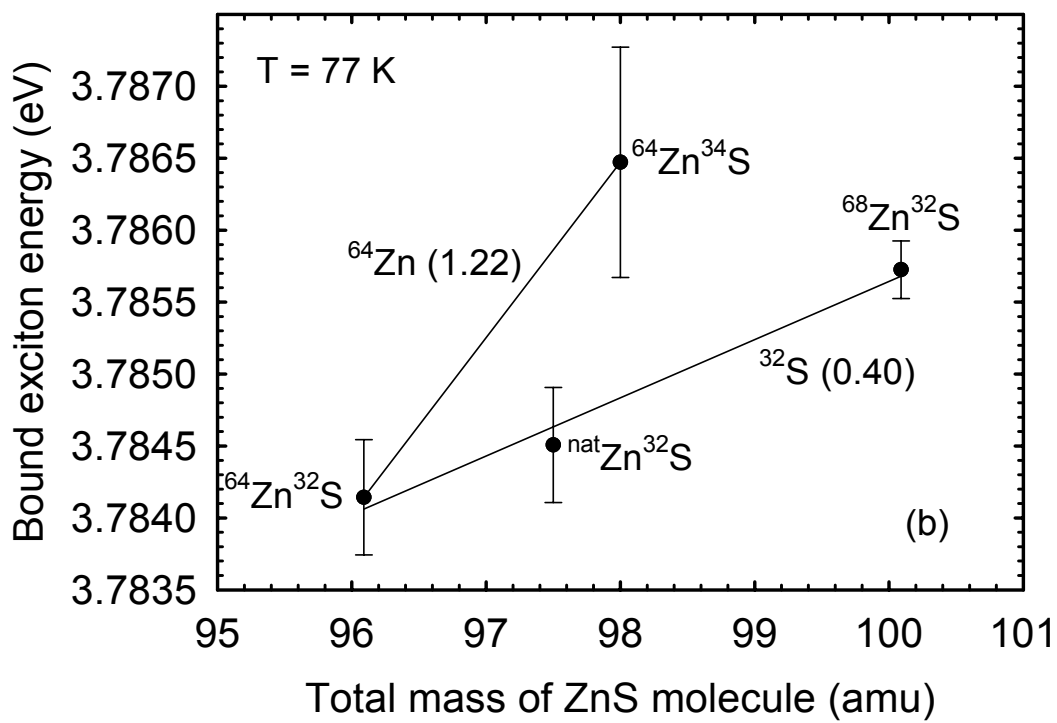

Fig.2.Manjón

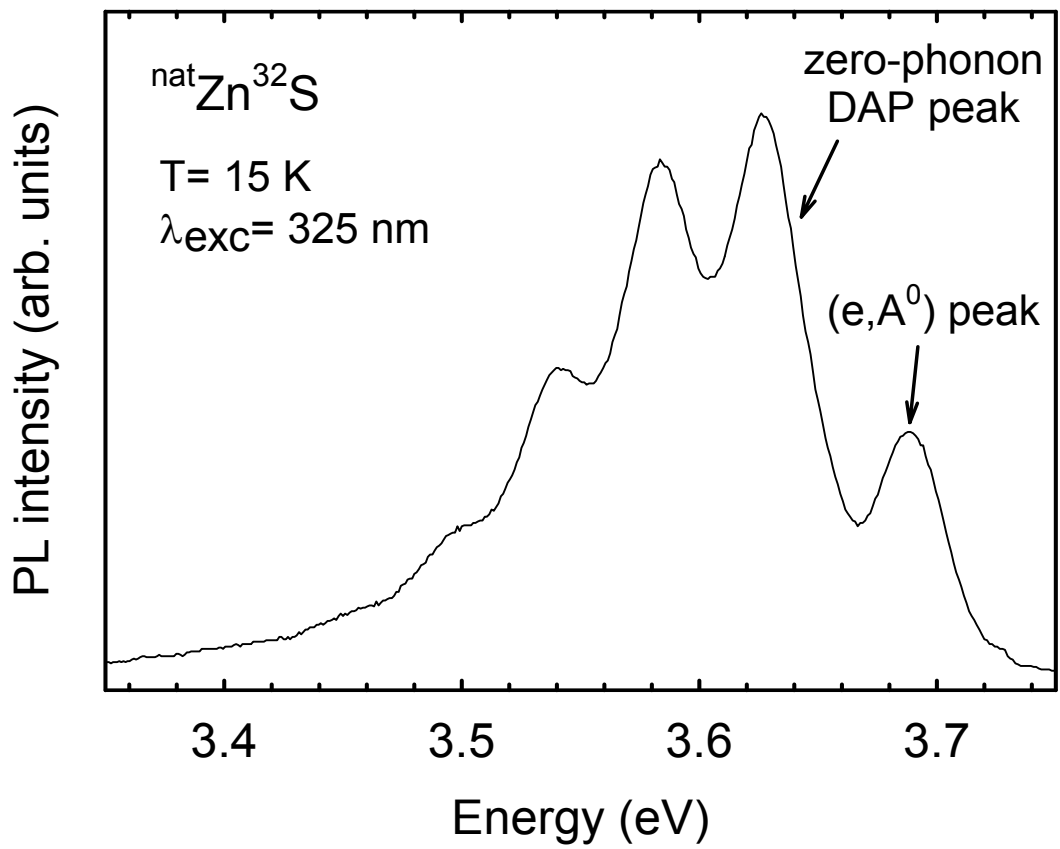

Fig. 3. Manjón

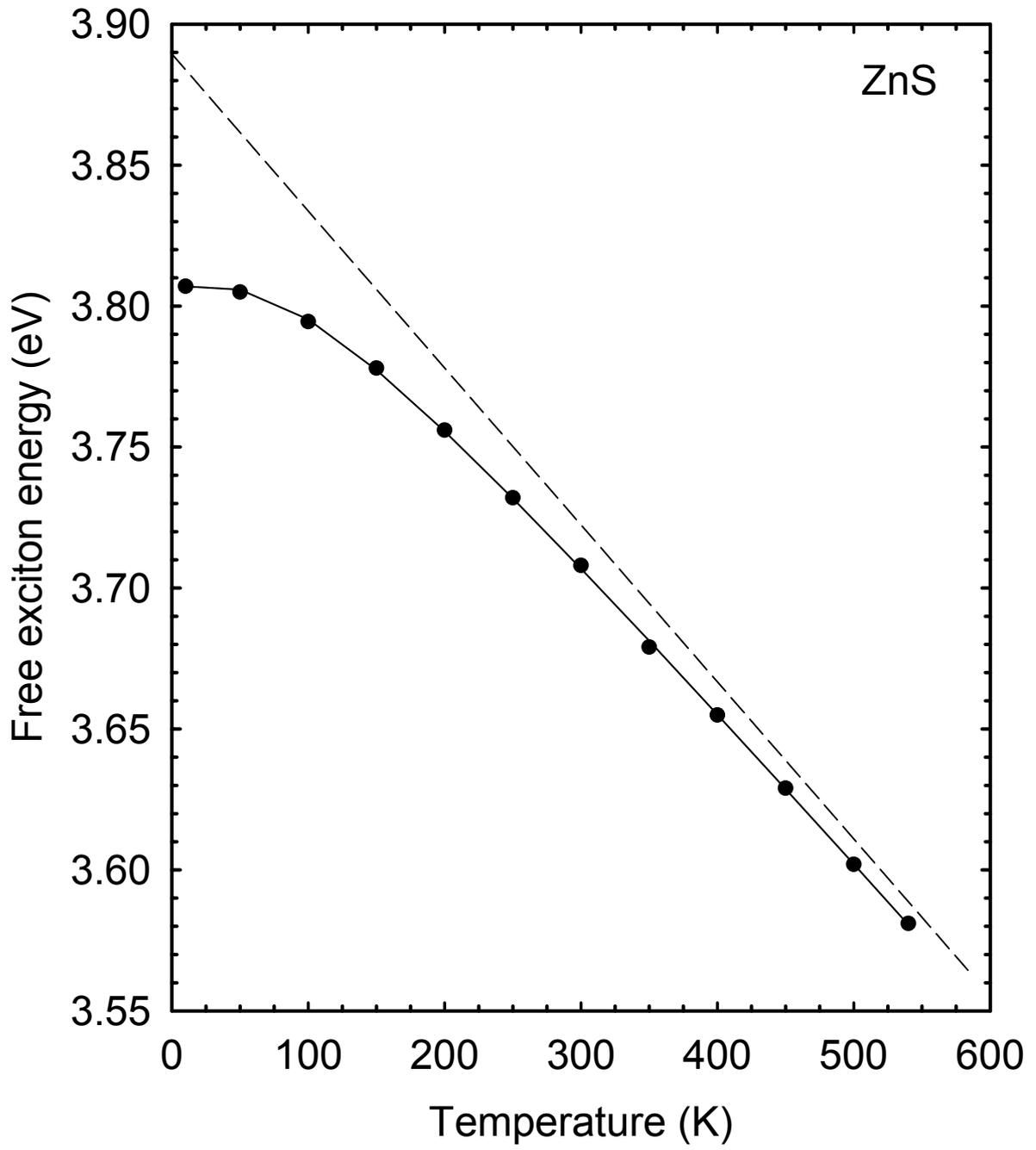

Fig. 4. Manjón